\documentclass[aps,prb,twocolumn]{revtex4} 

\usepackage{graphicx}
\usepackage{dcolumn}
\usepackage{bm}

\newcommand{\comment}[1]{}

\begin{document}
\renewcommand{\theequation}{\arabic{section}.\arabic{equation}}

\title{Quantum Monte Carlo in Classical Phase Space.
Mean Field and Exact Results for a One Dimensional Harmonic
Crystal.}


\author{Phil Attard}
\affiliation{{\tt phil.attard1@gmail.com}
\\ {\tt Projects/QSM19/mfxtl/mfxtl.tex} Begun 25-Mar-2019}


\begin{abstract}
Monte Carlo simulations are performed in classical phase space
for a one-dimensional quantum harmonic crystal.
Symmetrization effects for spinless bosons and fermions are quantified.
The algorithm is tested for a range of parameters against exact results
that use 20,000 energy levels.
It is shown that the singlet mean field approximation
is very accurate at high temperatures,
and that the pair  mean field approximation
gives a systematic improvement in the intermediate
and low temperature regime.
The latter is derived from
a cluster mean field approximation
that accounts for the non-commutativity of position and momentum,
and that can be applied in three dimensions.
\end{abstract}


\maketitle

%
\section{Introduction}
\setcounter{equation}{0} \setcounter{subsubsection}{0}
%

The impediment to applying quantum statistical mechanics
to condensed matter systems at terrestrial temperatures and densities
is the horrendous scaling with system size
entailed by conventional approaches.
\cite{Parr94,Morton05,Bloch08,McMahon12,Pollet12,Hernando13}
Although the field is not exactly moribund,
it would be fair to say that
the rate of progress has been disappointing,
and that there appears no obvious way of overcoming
this fundamental limitation of the existing methods.

The pressing need to approach the problem from a different direction
has motivated the author to develop a new methodology
that is based on a formally exact transformation
that expresses the quantum partition function
as an integral over classical phase space.\cite{QSM,STD2,Attard18a}
The validity of the formulation has been verified
analytically and numerically
for the quantum ideal gas,\cite{QSM,STD2}
and for non-interacting quantum harmonic oscillators.\cite{Attard18b}

Application of the algorithm to interacting systems
indicate that, for a given statistical error,
the algorithm
scales sub-linearly with system size.\cite{Attard16,Attard17,Attard18c}
This places it in a class of its own
for the treatment of condensed matter quantum systems.

Validation of the algorithm for an interacting system
has been provided by quantitative  tests
for interacting Lennard-Jones particles
against benchmarks obtained with more conventional methods
by Hernando and Van\'i\v cek.\cite{Hernando13}
The latter results were exact,
apart from the fact that they were numerical
and `only' 50 energy levels were used for the statistical averages.
One should not understate the difficulty of obtaining numerically
50 energy levels for the Lennard-Jones system;
it is undoubtedly a greater computational challenge
than obtaining analytically the 20,000 energy levels used below
for the present one-dimensional harmonic crystal.
If nothing else the figure of 50 levels does underscore
the insurmountable intractability of conventional quantum approaches.

In fact the original motivation for the present paper
was to sort out a small discrepancy between the putative exact results
of  Hernando and Van\'i\v cek\cite{Hernando13}
and the author's mean field, classical phase space results
at the highest temperature studied.\cite{Attard18c}
It was unclear whether the differences in the original comparison
were due to the mean field approximation used in the phase space simulation,
or else to the limited number of energy levels used in the exact results.
Accordingly the author has undertaken to establish
his own  exact results for use as benchmarks,
and it was found that 20,000 energy levels were necessary
for reliable results at temperatures high enough
to give a departure from the ground state.

The results of these tests are reported below.
It is found that the mean field approximation as originally formulated
is essentially exact at the highest temperatures,
but the error can be on the order of 5--10\%
at intermediate and low temperatures,
depending upon model parameters.
Accordingly, this papers develops a systematic general improvement
to the phase space algorithm
that might be called the cluster mean field approximation.
This is here implemented at the singlet level,
which is the original version,
and also at the pair level, which is new.
It is found that the pair  mean field approximation
gives essentially exact results at intermediate temperatures.
At low temperatures such that the system is predominantly
in the ground state,
the singlet and pair mean field approximations
are found to be in error by on the order of 5\%.

This paper relies upon earlier work,
which will not be re-derived here.
The reader is referred Ref.~\onlinecite{STD2}
for the derivation of quantum statistical mechanics,
to  Ref.~\onlinecite{Attard18a} for the derivation
of the phase space formulation
(see also  Ref.~\onlinecite{Attard19} for some formal details
concerning symmetrization of multiparticle states),
and to  Ref.~\onlinecite{Attard19} for the exact phonon analysis
of the present one-dimensional harmonic crystal.

%
\section{Analysis and Model}
\setcounter{equation}{0} \setcounter{subsubsection}{0}
%

\subsection{Phase Space Formulation}

The details of the classical phase space formulation
of quantum statistical mechanics can vary
with the particular quantity being averaged.\cite{Attard18a}
For the average energy,
which represents a certain class of operators,
namely those that are a linear combination
of functions purely of the momentum operator or of the position operator,
the canonical equilibrium average can be written\cite{Attard18a}
\begin{eqnarray} 
\lefteqn{
\left< \hat {\cal H} \right>_{N,V,T}^\pm
}  \\
& = &
\ \frac{1}{h^{dN} N!Z^\pm }
\int \mathrm{d}{\bf \Gamma}\;
e^{-\beta{\cal H}({\bf \Gamma})}
 W_p({\bf \Gamma})  \eta^\pm_q({\bf \Gamma})
 {\cal H}({\bf \Gamma}) , \nonumber
\end{eqnarray}
where the partition function is
\begin{eqnarray} 
\lefteqn{
Z^\pm(N,V,T)
}  \\
& = &
\ \frac{1}{h^{dN} N!  }
\int \mathrm{d}{\bf \Gamma}\;
e^{-\beta{\cal H}({\bf \Gamma})}
W_p({\bf \Gamma})  \eta^\pm_q({\bf \Gamma}) . \nonumber
\end{eqnarray}
In these $N$ is the number of particles,
assumed identical and spinless,
$V$ is the volume, $T$ is the temperature,
$\beta = 1/k_\mathrm{B}T$ is the inverse temperature,
$h$ is Planck's constant,
and $k_\mathrm{B}$ is Boltzmann's constant.
Also,
${\bf \Gamma} = \{ {\bf p},{\bf q} \}$ is a point in phase space,
with the vector of particles' momenta being
$ {\bf p} = \{ {\bf p}_1,{\bf p}_2, \ldots, {\bf p}_N \}$,
and that of the particles' position being
$ {\bf q} = \{ {\bf q}_1,{\bf q}_2, \ldots, {\bf q}_N \}$.
Finally, $\hat{\cal H} = {\cal H}(\hat{\bf p},\hat{\bf q})$
is the energy or Hamiltonian operator,
and ${\cal H}({\bf p},{\bf q})$ is the classical Hamiltonian function.

The plus sign is for bosons and the minus sign is for fermions.
Actually, since this has been formulated for spinless particles,
these refer to the fully symmetrized and fully anti-symmetrized
spatial part of the wave function
(see Appendix C of Ref.~\onlinecite{Attard19}).
Throughout the words `boson' and `fermion' should be understood in this sense.

The unsymmetrized position and momentum eigenfunctions
in the position representation
are
respectively\cite{Messiah61}
\begin{equation}
|{\bf q}\rangle = \delta({\bf r}-{\bf q})
, \mbox{ and }
|{\bf p}\rangle
=
\frac{e^{-{\bf p}\cdot{\bf r}/i\hbar}}{V^{N/2} } .
\end{equation}

The symmetrization function is defined as\cite{Attard18a}
\begin{equation}
\eta^\pm_q({\bf p},{\bf q})
\equiv
\frac{1}{\langle {\bf p} | {\bf q} \rangle }
\sum_{\hat{\mathrm P}} (\pm 1)^p \,
\langle \hat{\mathrm P} {\bf p} | {\bf q} \rangle .
\end{equation}
Here the sum is over the $N!$ permutation operators $\hat{\mathrm P}$,
whose parity is $p$.
The imaginary part of these is odd in momentum,
$\eta^\pm_q({\bf p},{\bf q})^* = \eta^\pm_q(-{\bf p},{\bf q})$.

The symmetrization function can be written as a series of loop products,
\begin{eqnarray}
\eta^\pm_q({\bf \Gamma})
& = &
1
+ \sum_{ij}\!'  \eta_{q;ij}^{\pm(2)}
+ \sum_{ijk}\!'  \eta_{q;ijk}^{\pm(3)}
\nonumber \\ &&  \mbox{ }
+ \sum_{ijkl}\!' \eta_{q;ij}^{\pm(2)}
\eta_{q;kl}^{\pm(2)}
+ \ldots
\end{eqnarray}
Here the superscript is the order of the loop,
and the subscripts are the atoms involved in the loop.
The prime signifies that the sum is over unique loops
(ie.\ each configuration of particles in loops occurs once only)
with each index different
(ie.\ no particle may belong to more than one loop).
In general
the $l$-loop symmetrization factor is
\begin{equation} \label{Eq:tilde-eta-l}
\eta_{q;1 \ldots l}^{\pm(l)}
=
(\pm 1)^{l-1}
e^{ {\bf q}_{1l} \cdot {\bf p}_l /i\hbar }
\prod_{j=1}^{l-1}
e^{ {\bf q}_{j+1,j} \cdot {\bf p}_j /i\hbar } ,
\end{equation}
where $ {\bf q}_{jk} \equiv {\bf q}_{j}- {\bf q}_{k}$.
This corrects a typographical error
in Eq.~(3.4) of Ref.~\onlinecite{Attard18a}

The commutation function,
which is essentially the same as the function
introduced by Wigner\cite{Wigner32}
and analyzed by Kirkwood,\cite{Kirkwood33}
is defined by
\begin{equation}\label{Eq:def-Wp}
e^{-\beta{\cal H}({\bf p},{\bf q})}
W_{p}({\bf p},{\bf q})
=
\frac{\langle{\bf q}|e^{-\beta\hat{\cal H}} |{\bf p}\rangle
}{\langle{\bf q}| {\bf p}\rangle } .
\end{equation}
Again one has
$W_p({\bf p},{\bf q})^* =
W_p(-{\bf p},{\bf q})$.
High temperature expansions for the commutation function
have been given.\cite{Wigner32,Kirkwood33,STD2,Attard18b}

The commutation function in phase space can also be written
as a series of energy eigenfunctions and eigenvalues,
$\hat{\cal H} |{\bf n}\rangle = E_n  |{\bf n}\rangle$.
Using the completeness properties of these
one obtains
\begin{eqnarray}
e^{-\beta{\cal H}({\bf p},{\bf q})}
W_{p}({\bf p},{\bf q})
& = &
\frac{\langle{\bf q}|e^{-\beta\hat{\cal H}} |{\bf p}\rangle
}{\langle{\bf q}| {\bf p}\rangle }
 \\ & = &
\frac{1}{\langle{\bf q}| {\bf p}\rangle }
\sum_{\bf n}
e^{-\beta E_{\bf n}} \,
\langle{\bf q}|{\bf n}\rangle \,
\langle{\bf n} |{\bf p}\rangle .\nonumber
\end{eqnarray}
This exact expression
forms the basis of the mean field approximation
to the commutation function.

\subsection{Cluster Mean Field Approximation}

In the classical phase space formulation of quantum statistical mechanics,
the symmetrization function is relatively trivial to obtain and implement.
The commutation function is more of a challenge,
with the most successful approach using a mean field approximation
that exploits the analytic form
of the commutation function in the case of independent simple
harmonic oscillators.\cite{Attard18b}
This has previously been tested for the simulation of a Lennard-Jones system.
\cite{Attard18c}

This section begins with a summary of
the singlet  mean field approximation
to the commutation function.\cite{Attard18b}
Then the cluster mean field approximation is given.

\subsubsection{Singlet Mean Field Approximation}


In general, the particles of the sub-system
interact via the potential energy,
which is the sum of one-body,
two-body, three-body terms, etc.,
\begin{eqnarray}
U({\bf q}) & = &
\sum_{j=1}^N u^{(1)}({\bf q}_j)
+ \sum_{j<k}^N u^{(2)}({\bf q}_j,{\bf q}_k)
\nonumber \\ && \mbox{ }
+ \sum_{j<k<l}^N u^{(3)}({\bf q}_j,{\bf q}_k,{\bf q}_l)
+ \ldots
\end{eqnarray}
Distributing the energy equally,
the energy of particle $j$ 
can be defined as
\begin{eqnarray}
U_j({\bf q}_j;{\bf q})
& = &
u^{(1)}({\bf q}_j)
+ \frac{1}{2} \sum_{k=1}^N\!^{(k\ne j)} \, u^{(2)}({\bf q}_j,{\bf q}_k)
\nonumber \\ && \mbox{ }
+ \frac{1}{3}
\sum_{k<l}^N\!^{(k,l\ne j)} \, u^{(3)}({\bf q}_j,{\bf q}_k,{\bf q}_l)
+ \ldots
\end{eqnarray}
with
$ U({\bf q}) = \sum_{j=1}^N U_j({\bf q}_j;{\bf q})$.
The argument $({\bf q}_j;{\bf q})$ means that ${\bf q}_j$
is here separated out from ${\bf q}$.

The potential energy of particle $j$ in configuration ${\bf q}$
may be expanded to second order
about its local minimum at $\overline{\bf q}_j({\bf q})$,
\begin{equation}
U_j({\bf q}_j;{\bf q})
=
\overline U_j({\bf q})
+ \frac{1}{2} [{\bf q}_j-\overline{\bf q}_j][{\bf q}_j-\overline{\bf q}_j]
: \overline{\underline{\underline U}}_j'' ,
\end{equation}
where the minimum  value of the potential is
$\overline U_j({\bf q}) \equiv U_j(\overline{\bf q}_j;{\bf q})$.
The  $d \times d$ second derivative matrix for particle $j$
at the minimum,
$\overline{\underline{\underline U}}_j''
= \left. \nabla_j \nabla_j  U_j({\bf q}_j;{\bf q})
 \right|_{{\bf q}_j = \overline{\bf q}_j}$,
is assumed positive definite.

For configurations ${\bf q}$
that have no local minimum in the potential,
or that have too large a displacement $|{\bf q}_j- \overline{\bf q}_j|$,
the corresponding single particle commutation function
can be set to unity, $W_j({\bf \Gamma}) = 1$,
(or, in the multi-dimensional case,
the commutation function of the corresponding mode).
This is justified by analytic results
for the simple harmonic oscillator.\cite{Attard18b}

The positive definite second derivative matrix
has $d$ eigenvalues $\lambda_{j\alpha}({\bf q})  > 0$,
and  orthonormal eigenvectors,
$ \overline{\underline{\underline U}}_j'' {\bf X}_{j\alpha}
= \lambda_{j\alpha}{\bf X}_{j\alpha}$, $\alpha = x,y,\ldots, d$.
For molecule $j$ in configuration ${\bf q}$ the eigenvalues
define the frequencies
\begin{equation}
\omega_{j\alpha}({\bf q}) = \sqrt{ \lambda_{j\alpha}({\bf q}) /m }
, \;\; \alpha = x, y, \ldots , d .
\end{equation}
With this the potential energy is
\begin{eqnarray}
U({\bf q})
& = &
\sum_{j=1}^N \overline U_j
+
 \frac{1}{2}\sum_{j=1}^N
({\bf q}_j-\overline{\bf q}_j) ({\bf q}_j-\overline{\bf q}_j) :
\overline{\underline{\underline U}}_j''
\nonumber \\ & = &
\sum_{j=1}^N  \overline U_j
+
\frac{1}{2} \sum_{j,\alpha}  \hbar \omega_{j\alpha} Q_{j\alpha}^2 .
\end{eqnarray}
Here $ Q_{j\alpha} \equiv
\sqrt{ { m \omega_{j\alpha} }/{\hbar} } \, Q_{j\alpha}' $,
and
$ {\bf Q}_j' =
 \underline{\underline X}_j^\mathrm{T}  [{\bf q}_j-\overline{\bf q}_j] $.
(This corrects a typographical error in Eqs~(2.7) and (2.8)
in Ref.~\onlinecite{Attard18c}.)


The mean field approximation combined with
the second order expansion about the local minima
maps each configuration ${\bf \Gamma}$
to a system of $dN$ independent harmonic oscillators
with frequencies $\omega_{j\alpha}$
displacements $Q_{j\alpha}$,
and momenta $P_{j\alpha} =
\left\{ \underline{\underline X}_j^\mathrm{T}
{\bf p}_j \right\}_\alpha / \sqrt{ m\hbar\omega_{j\alpha} }$.
(This corrects a typographical error in Ref.~\onlinecite{Attard19}.)

With this harmonic approximation
for the potential energy,
the effective Hamiltonian in a particular configuration can be written
\begin{equation} \label{Eq:H^SHO}
{\cal H}^\mathrm{SHO}({\bf p},{\bf q}-\overline{\bf q})
=
\sum_{j=1}^N \overline U_j
+
\frac{1}{2} \sum_{j,\alpha}  \hbar \omega_{j\alpha}
\left[ P_{j\alpha}^2 + Q_{j\alpha}^2 \right] .
\end{equation}
The commutation function for the interacting system
for a particular configuration
can be approximated as the product of commutation functions
for effective non-interacting harmonic oscillators
which have the local displacement as their argument.
With this the mean field commutation function is
\begin{eqnarray} \label{Eq:Wmf}
W_p^\mathrm{mf}({\bf \Gamma})
& \approx &
W_p^\mathrm{SHO}({\bf p},{\bf q}-\overline{\bf q})
\nonumber \\ & = &
e^{\beta {\cal H}^\mathrm{SHO}({\bf p},{\bf q}-\overline{\bf q})}
\frac{ \langle {\bf q}-\overline{\bf q}
|e^{-\beta \hat{\cal H}^\mathrm{SHO} }
|{\bf p}\rangle
}{ \langle {\bf q}-\overline{\bf q}|{\bf p}\rangle }
\nonumber \\ & = &
\prod_{j,\alpha} W_{p,j\alpha}^\mathrm{SHO}(P_{j\alpha},Q_{j\alpha}) .
\end{eqnarray}
The harmonic oscillator commutation function
for a single mode is\cite{Attard18b}
\begin{eqnarray} \label{Eq:WSHO}
\lefteqn{
W_{p,{j\alpha}}^\mathrm{SHO}(P_{j\alpha},Q_{j\alpha})
}  \\
& = &
\sqrt{2}
e^{-iP_{j\alpha}Q_{j\alpha}}
e^{\beta\hbar \omega_{j\alpha} [P_{j\alpha}^2+Q_{j\alpha}^2]/2 }
e^{-[P_{j\alpha}^2+Q_{j\alpha}^2]/2 }
\nonumber \\ && \mbox{ } \times
\sum_{n_{j\alpha}=0}^\infty
\frac{i^{n_{j\alpha}}
e^{- \beta \hbar \omega_{j\alpha}  (n_{j\alpha}+1/2) }
}{
2^{n_{j\alpha}} {n_{j\alpha}}!  }
\mathrm{H}_{n_{j\alpha}}(P_{j\alpha})
\mathrm{H}_{n_{j\alpha}}(Q_{j\alpha}) .\nonumber
\end{eqnarray}
The prefactor $e^{-iP_{j\alpha}Q_{j\alpha}}$
corrects the prefactor $e^{-ip_{j\alpha}q_{j\alpha}/\hbar}$
given in  Eq.~(5.10) of Ref.~[\onlinecite{Attard18b}].
Here $\mathrm{H}_n(z)$ is the Hermite polynomial of degree $n$.
The imaginary terms here are odd in momentum.
As justified by analytic results
for the simple harmonic oscillator,\cite{Attard18b}
for configurations
such that $\overline{\underline{\underline U}}_j''({\bf q})$
is not positive definite
(ie.\ a particular eigenvalue is not positive, $\lambda_{j\alpha} \le 0$),
or that the displacement $Q_{j\alpha}$ exceeds a predetermined cut-off,
the corresponding commutation function can be set to unity,
$W_{p,{j\alpha}}^\mathrm{SHO} = 1$.

For the averages, the momentum integrals
can be performed analytically,
both here and in combination with the symmetrization function.
This considerably reduces computer time
and substantially increases accuracy.

\subsubsection{Cluster Mean Field Approximation}

Any configuration ${\bf q}$
can be decomposed into disjoint clusters labeled $\alpha=1,2,\ldots$.
Of the different criteria that can be used to define a cluster,
perhaps the simplest is that two particles  belong to the same
cluster if, and only if,
they are connected by at least one path of bonds.
Two particles are bonded
if their separation is less than a nominated  length.
Some clusters, perhaps the great majority,
will consist of a single particle.

An even simpler definition can be made in one dimension.
In this case define the pair cluster $\alpha$,
$\alpha =1,2,\ldots, N/2$,
as the nearest neighbors $\{2\alpha-1, 2\alpha\}$,
irrespective of their actual separation.
This criteria is used in the results presented below.

Using a separation-based criterion for the definition of a cluster
is useful not only for the mean field approximation
to the commutation function,
but also for the calculation of the symmetrization function.
Depending on the chosen bond length,
only permutations of particles within the same cluster
need to be considered.
(This idea is not used in the results presented below.)

The cluster energy is the internal energy
plus the relevant proportion of the interaction energy with other clusters:
half for pair interactions,
one third for triplet interactions, etc.
The total potential energy is
\begin{eqnarray}
U({\bf q}) & = &
\sum_{j=1}^N u^{(1)}({\bf q}_j)
+
\sum_{j<k} u^{(2)}({\bf q}_j,{\bf q}_k)
+ \ldots
\nonumber \\ & = &
\sum_\alpha U_\alpha({\bf q}_\alpha;{\bf q}) ,  
\end{eqnarray}
where the  energy of cluster $\alpha$ is
\begin{eqnarray}
U_\alpha({\bf q}_\alpha;{\bf q})  
& = &
\sum_{j\in \alpha}  u^{(1)}({\bf q}_j)
+
\sum_{j<k}\!^{(j,k \in \alpha)}  u^{(2)}({\bf q}_j,{\bf q}_k)
\nonumber \\ && \mbox{ }
+\frac{1}{2}
\sum_{j\in \alpha}
\sum_\beta \!^{(\beta \ne \alpha)} \sum_{k\in \beta}
u^{(2)}({\bf q}_j,{\bf q}_k)
\nonumber \\ && \mbox{ }
+ \ldots .
\end{eqnarray}
There are $N_\alpha$ particles in cluster $\alpha$,
with positions ${\bf q}_\alpha =
\{  {\bf q}_{\alpha,1}, {\bf q}_{\alpha,2}, \ldots,
{\bf q}_{\alpha,N_\alpha} \}$,
where $ {\bf q}_{\alpha,j} =  {\bf q}_{k}$ for one of the $k \in \alpha$.
This is a $(d N_\alpha)$-dimensional vector.

The second order expansion about
the minimum energy cluster configuration, $\overline {\bf q}_\alpha$,
is
\begin{equation}
U_\alpha({\bf q}_\alpha;{\bf q})  
=
\overline U_\alpha({\bf q})  
+ \frac{1}{2}
\overline{\underline{\underline U}}_\alpha'' :
[{\bf q}_\alpha- \overline{\bf q}_\alpha ]
[{\bf q}_\alpha- \overline{\bf q}_\alpha ] .
\end{equation}
The second derivative matrix is
$ \overline{\underline{\underline U}}_\alpha''
\equiv
\left.
\nabla_\alpha \nabla_\alpha U_\alpha({\bf q}_\alpha;{\bf q})
\right|_{\overline {\bf q}_\alpha}$,
which is $(d N_\alpha)\times (d N_\alpha)$.
One has to find the eigenvalues, assumed positive,
and eigenvectors of this.
These give the cluster phonon mode frequencies $\omega_{\alpha,b}$,
$b=1,2,\ldots,d N_\alpha$,
and mode amplitudes,
${\bf Q}_\alpha  =
\underline{\underline \omega}_\alpha
\underline{\underline X}_\alpha^\mathrm{T}
[{\bf q}_\alpha- \overline{\bf q}_\alpha ]$.
The frequency matrix is diagonal with elements
$\omega_{\alpha;bc} = \sqrt{m \omega_{\alpha,b}/\hbar}\, \delta_{bc}$.
As before, the momentum of mode $b$ in cluster $\alpha$ is
$P_{\alpha,b} =
\left\{ \underline{\underline X}_\alpha^\mathrm{T}
{\bf p}_{\alpha} \right\}_b / \sqrt{ m\hbar\omega_{\alpha,b} }$.

This formulation is essentially the same as the singlet
mean field theory,
and one may similarly define the cluster mean field
commutation function as the product of simple harmonic oscillator
commutation functions, one for each phonon mode of each cluster.
%
The cluster mean field commutation function of the configuration is
\begin{equation}
W^\mathrm{mf}({\bf \Gamma})
=
\prod_\alpha
W_\alpha^\mathrm{SHO}({\bf P}_\alpha,{\bf Q}_\alpha)
=
\prod_{\alpha,b}
W_{\alpha,b}^\mathrm{SHO}({P}_{\alpha,b},{Q}_{\alpha,b}).
\end{equation}

From the computational point of view,
a felicitous aspect of the cluster mean field approximation
is that there are exactly as many modes
as in the singlet mean field approximation.
This means that all of the sub-routines called
to obtain the various statistical averages in the singlet approximation
can be called without change in the cluster mean field approximation.

\subsection{Harmonic Crystal Model}

\subsubsection{Potential Energy}

Following earlier work,\cite{TDSM,Attard19}
consider a one-dimensional harmonic crystal in which the particles
are attached by linear springs to each other and to lattice sites.
Let the coordinate of the $j$th particle
be $q_{j}$, and let its lattice position
(ie.\ in the lowest energy state) be $\tilde{\overline q}_{j} = j \Delta_q$.
The tilde signifies that these are ordered.
The lattice spacing is also the relaxed inter-particle spring length.
There are fixed `wall' particles at $q_{0} = 0$
and $q_{N+1} = (N+1)\Delta_q$.
Let $d_j \equiv q_{j}- \tilde{\overline q}_{j}$
be the displacement from the lattice position;
for the wall particles, $d_0=d_{N+1}=0$.
The system has over-all number density $\rho = \Delta_q^{-1}$.

There is an external harmonic potential
of spring constant $\kappa$
acting on each particle centered at its lattice site.
The inter-particle spring has strength $\lambda$
and relaxed length $\Delta_q$.
With these the potential energy is
\begin{eqnarray}  \label{Eq:U(q)}
U({\bf q})
& = &
\frac{\kappa}{2} \sum_{j=1}^N [q_{j}-\tilde{\overline q}_{j} ]^2
+
\frac{\lambda}{2} \sum_{j=0}^N [q_{j+1}-q_{j}-\Delta_q]^2
\nonumber \\ & = &
\frac{\kappa}{2} \sum_{j=1}^N d_j^2
+
\frac{\lambda}{2} \sum_{j=0}^N [d_{j+1}-d_j]^2.
\end{eqnarray}
The energy eigenfunctions and eigenvectors
can be obtained explicitly for this model
by expressing it in terms of phonon modes.\cite{Attard19}

This model potential is \emph{not} invariant
with respect to the permutation of the positions of the particles,
and it therefore violates a fundamental axiom of quantum mechanics.
\cite{Messiah61}
This was discussed in detail in Appendix A of Ref.~\onlinecite{Attard19},
where the origin, interpretation, and justification for the potential
was given.
To those remarks may be added the fact
that the formulation of quantum statistical mechanics
in classical phase space is unchanged by a non-symmetric potential
(unpublished).

\subsubsection{Singlet Mean Field}

The energy of particle $j$ in configuration ${\bf q}$ is
\begin{eqnarray}
\lefteqn{
U_j(q_j;{\bf q})
} \nonumber \\
& = &
\frac{\kappa}{2} [q_{j}-\tilde{\overline q}_{j} ]^2
+
\frac{\lambda}{4} [q_{j}-q_{j+1}+\Delta_q]^2
\nonumber \\ && \mbox{ }
+
\frac{\lambda}{4} [q_{j}-q_{j-1}-\Delta_q]^2
\nonumber \\ && \mbox{ }
+ \frac{\lambda}{4} [q_{1}-\Delta_q]^2 \delta_{j1}
+ \frac{\lambda}{4} [q_{N}-N\Delta_q]^2 \delta_{jN}
\nonumber \\ & = &
\frac{\kappa}{2} d_j^2
+
\frac{\lambda}{4}
\left\{ [d_{j}-d_{j+1}]^2 + [d_{j}-d_{j-1}]^2 \right\}
\nonumber \\ && \mbox{ }
+
\frac{\lambda}{4} d_{j}^2 [\delta_{j1} + \delta_{jN} ].
\end{eqnarray}
The total potential energy is just
$ U({\bf q}) = \sum_{j=1}^N U_j({\bf q}_j;{\bf q})  $,
and $d_j \equiv q_{j}-\tilde{\overline q}_{j}$.

The gradient vanishes when
\begin{eqnarray}
\overline d_j({\bf q})
& = &
\frac{\lambda }{2 \overline U''_j} [d_{j+1} + d_{j-1} ] .
\end{eqnarray}
The second derivative is
\begin{equation}
\overline U''_j
\equiv
\nabla_j \nabla_j  U_j(q_j;{\bf q})
=
\kappa + \lambda + \frac{\lambda}{2}[\delta_{j1} + \delta_{jN} ] ,
\end{equation}
which is independent of the configuration ${\bf q}$.

The potential energy of particle $j$ in configuration ${\bf q}$
may be expanded to second order
about its local minimum at $\overline{q}_j({\bf q})$,
\begin{eqnarray}
U_j({q}_j;{\bf q})
& = &
\overline U_j({\bf q})
+ \frac{\overline U''_j}{2}
[{q}_j-\overline{q}_j({\bf q})]^2 .
\end{eqnarray}
where 
$\overline U_j({\bf q}) \equiv U_j(\overline{q}_j({\bf q});{\bf q})$.
This second order expansion for the potential is exact
for the present harmonic crystal.
Note that the most likely position of particle $j$
for the current configuration, $\overline{q}_j({\bf q})$,
is not the same as its lattice position $\tilde{\overline q}_{j} $.


For each molecule define the frequency
$ \omega_j  = \sqrt{ \overline U''_j /m } $.
This is the same for all configurations ${\bf q}$.
With this the potential energy is
\begin{eqnarray}
U({\bf q})
& = &
\sum_{j=1}^N \overline U_j
+ \frac{1}{2}  \sum_{j=1}^N
\overline U''_j  [{q}_j-\overline{q}_j({\bf q})]^2
\nonumber \\ & = &
\sum_{j=1}^N  \overline U_j
+
\frac{1}{2} \sum_{j}  \hbar \omega_{j} Q_{j}^2 .
\end{eqnarray}
Here $ Q_{j} \equiv
\sqrt{ { m \omega_{j} }/{\hbar} } \, [ {q}_j-\overline{ q}_j] $.
As above, the momenta are
${P}_{j} = {p}_j/\sqrt{m\hbar \omega_{j}}  $.

This is the approximation in the singlet mean field approach:
each frequency mode is a displacement of a single particle.
One may now apply
Eqs~(\ref{Eq:H^SHO}), (\ref{Eq:Wmf}), and (\ref{Eq:WSHO})
for the singlet mean field commutation function.

\subsubsection{Pair Mean Field Approximation}

For the one dimensional harmonic crystal,
define a cluster pair $\alpha$
as the nearest neighbors $\{2\alpha-1, 2\alpha\}$,
 $\alpha=1,2,\ldots, \lfloor N/2 \rfloor$.
For simplicity, assume $N$ even.
(Contrariwise, include particle $N$ as a singlet cluster.)

The energy of cluster $\alpha$ is
\begin{eqnarray}
\lefteqn{
U_\alpha({\bf q}_\alpha;{\bf q})
}  \\
& = &
\frac{\kappa}{2}
\left\{ d_{2\alpha-1}^2 + d_{2\alpha}^2 \right\}
+
\frac{\lambda}{2}
\left\{ \rule{0cm}{.4cm}
[d_{2\alpha} - d_{2\alpha-1}]^2
\right. \nonumber \\ && \mbox{ } \left.
+ \frac{1}{2} [d_{2\alpha-1} - d_{2\alpha-2}]^2
+ \frac{1}{2} [d_{2\alpha+1} - d_{2\alpha}]^2
\right\}
\nonumber \\ & &
+\frac{\lambda}{4} [d_{2\alpha-1} - d_{2\alpha-2}]^2 \delta_{2\alpha-1,1}
+  \frac{\lambda}{4} [d_{2\alpha+1} - d_{2\alpha}]^2 \delta_{2\alpha,N},
\nonumber
\end{eqnarray}
where the displacement is $d_j = q_j - \tilde{\overline q}_j$.
Note that the interaction with the wall particles,
when present,
has to be counted fully.
Note also that $d_0 = d_{N+1} = 0$.

The second derivative matrix is
\begin{eqnarray}
\overline{\underline{\underline U}}_\alpha''
& = &
\left( \begin{array}{cc}
\kappa + \frac{3 \lambda}{2} + \frac{\lambda}{2} \delta_{2\alpha-1,1}  &
 -\lambda  \\
-\lambda &
\kappa + \frac{3 \lambda}{2} + \frac{\lambda}{2} \delta_{2\alpha,N}
\end{array} \right)
\nonumber \\ & \equiv &
-\lambda
\left( \begin{array}{cc}
K_\alpha' & 1  \\ 1 & K_\alpha''
\end{array} \right) .
\end{eqnarray}
Using it gives the optimum cluster displacement as
\begin{equation}
\left( \begin{array}{c}
\overline d_{2\alpha-1} \\ \overline d_{2\alpha}
\end{array} \right)
=
\frac{-1/2}{K_\alpha'K_\alpha''-1}
\left( \begin{array}{c}
K_\alpha'' d_{2\alpha-2} -  d_{2\alpha+1} \\
- d_{2\alpha-2} + K_\alpha' d_{2\alpha+1}
\end{array} \right) .
\end{equation}

The eigenvalues of the second derivative matrix (without $-\lambda$) are
\begin{eqnarray}
\mu_\alpha^\pm
& = &
\frac{1}{2}( K_\alpha'+K_\alpha'')
\pm
\frac{1}{2}
\sqrt{( K_\alpha'-K_\alpha'')^2 + 4  } .
\end{eqnarray}
Since the $K_\alpha$ are negative, so are the $\mu_\alpha^\pm$.

Writing the eigenvectors as
${\bf u}_\alpha^\pm = c_\alpha^\pm  (1,a_\alpha^\pm )$,
$c_\alpha^\pm  \equiv 1/\sqrt{1+a_\alpha^{\pm2}}$,
from the eigenvalue equation one obtains
$a_\alpha^\pm  = \mu_\alpha^\pm - K_\alpha'$.
This is presumably equivalent to
 $a_\alpha^\mp  = 1/[ \mu_\alpha^\mp - K_\alpha'']$.

The mode frequencies are
$\omega_\alpha^\pm  = \sqrt{-\lambda \mu_\alpha^\pm /m}$,
which are real.
The mode amplitude is
\begin{eqnarray}
{\bf Q}_\alpha  & = &
\underline{\underline \omega}_\alpha
\underline{\underline X}_\alpha^\mathrm{T}
\Delta {\bf d}_\alpha
 \\ & = &
\left( \begin{array}{cc}
\sqrt{m \omega_\alpha^+ /\hbar} \left\{
c_\alpha^+  \Delta d_{2\alpha-1}
+ c_\alpha^+  a_\alpha^+  \Delta d_{2\alpha} \right\} \\
\sqrt{m \omega_\alpha^- /\hbar}\left\{
c_\alpha^- \Delta d_{2\alpha-1}
+ c_\alpha^- a_\alpha^-  \Delta d_{2\alpha} \right\}
\end{array} \right),\nonumber
\end{eqnarray}
where $ \Delta d_{2\alpha-1} \equiv  d_{2\alpha-1} - \overline d_{2\alpha-1}
= q_{2\alpha-1} - \overline q_{2\alpha-1}$,
and
$ \Delta d_{2\alpha} \equiv  d_{2\alpha} - \overline d_{2\alpha}
= q_{2\alpha} - \overline q_{2\alpha}$.

The contribution to the total energy from cluster $\alpha$
in configuration ${\bf \Gamma}$ is
\begin{eqnarray}
{\cal H}_\alpha({\bf \Gamma})
& = &
\frac{1}{2m} \left[ p_{2\alpha-1}^2 + p_{2\alpha}^2 \right]
\nonumber \\ && \mbox{ }
+
\overline U_\alpha
+
\frac{1}{2}
\overline{\underline{\underline U}}''_\alpha :
\left[ {\bf q}_{\alpha} - \overline {\bf q}_{\alpha}  \right]
\left[ {\bf q}_{\alpha} - \overline {\bf q}_{\alpha}  \right]
\nonumber \\ & = &
\overline U_\alpha
+ \frac{\hbar \omega_{\alpha}^{+}}{2}
\left\{ P_{\alpha,+}^2 + Q_{\alpha,+}^2  \right\}
\nonumber \\ && \mbox{ }
+ \frac{\hbar \omega_{\alpha}^{-}}{2}
\left\{ P_{\alpha,-}^2 + Q_{\alpha,-}^2  \right\} .
\end{eqnarray}
In the computational implementation of the algorithm,
these can be used  to replace directly the singlet mean field terms,
$P_{\alpha}^{+} \Rightarrow P_{2\alpha-1} $,
$P_{\alpha}^{-} \Rightarrow P_{2\alpha} $,
and similarly for $Q_{\alpha}^{\pm}$ and $\omega_{\alpha}^{\pm}$.

\subsubsection{Symmetrization Function}

Symmetrization consists of a sum over all particle permutations.
However because of the highly oscillatory Fourier contributions
to the loop symmetrization function, Eq.~(\ref{Eq:tilde-eta-l}),
only permutations of closely separated particles
actually contribute to the statistical average.

In view of this one can define the permutation length,
\cite{Attard19}
\begin{equation}
d_{\mathrm m}(\hat{\mathrm P} ) \equiv
\sum_{j=1}^{N} | j - j' |
, \;\;
j' \equiv \{ \hat{\mathrm P} {\bf j} \}_j .
\end{equation}
One can see that  $d_{\mathrm m} = 0$ corresponds to the identity permutation,
$d_{\mathrm m}=2$ corresponds to a single nearest neighbor transposition
(dimer),
and $d_{\mathrm m}=4$
corresponds to either two distinct nearest neighbor transpositions
(double dimer),
or else a single cyclic permutation of three consecutive particles
(trimer).
One expects that the contributions to the symmetrized wave function
will decrease with increasing permutation length.

Hence one can set an upper limit on the length of the permutations
that are included in the symmetrization function.
The numerical results below show that by far the greatest contribution
comes from the identity permutation alone, $d_{\mathrm m}=0$.
In some cases a measurable change occurs by including also
nearest neighbor permutations, $d_{\mathrm m}=2$.
Measurable but smaller change occurs
upon also including permutations of length $d_{\mathrm m}=4$
(not shown below).
For $N$ particles,
the number of  permutation terms that contribute
to the symmetrization function is 1 for $d_{\mathrm m}  =0$,
$1 + (N-1)$ for $d_{\mathrm m} \le 2$,
and  $N + (N-2)(N-3)/2 + 2(N-2) $ for $d_{\mathrm m} \le 4$.

\subsection{Simulation Algorithm}

The simulation algorithm was as previously described.\cite{Attard18c}
Briefly the Metropolis algorithm in position space was used
with the usual classical Maxwell-Boltzmann weight.
The various momentum integrals were performed analytically.
Averages were evaluated by umbrella sampling
using the commutation function and symmetrization function as weight.
Since three versions of the commutation function
(unity (ie.\ classical), singlet mean field, pair mean field),
and three versions of the symmetrization function
($d_\mathrm{m}=0$ (ie.\ classical),
and $d_\mathrm{m}\le 2$, for bosons and for fermions)
were tested,
9 different averages for each quantity were obtained simultaneously.

Typically enough configurations were generated to make
the statistical error less than 1\%,
sometimes much less.
In the simulations the time depends on how many
Hermite polynomials are used for the commutation function
(eight in the results reported below; tests with six and twelve
showed no great effect),
and the cut-off for the mode amplitude beyond which the commutation
function was set to unity ($Q^\mathrm{cut} = 1$ in the results below;
tests with $Q^\mathrm{cut} = 2$ showed no great effect).

Interestingly enough,
for an accuracy of about 1\%,
the Monte Carlo algorithm was a factor of about $2,000$ times more
efficient (in terms of total computer time)
than the quasi-analytic exact phonon method.\cite{Attard19}
The main bottleneck in the latter
was the crude numerical quadrature method
that was used to evaluate the symmetrization function and density profile,
and this was exacerbated by the large number of energy levels
that were required for accurate results at higher temperatures.
For this particular comparison at $\beta \hbar \omega_\mathrm{LJ}=2$,
$l^\mathrm{max}=20,000$ energy levels were necessary;
grid parameters $\Delta_Q = 0.15 $ and $Q^\mathrm{max} = 6$
gave a quadrature error of about .8\%.
The phonon method is more efficient in one respect,
namely that it requires negligible  computer time
for each additional temperature point;
the simulations give results for only one temperature at a time.

The Lennard-Jones frequency used to scale the results below is
$\omega_\mathrm{LJ} = 3.28 \times 10^{12}\,$Hz,
the mass is $m=3.35 \times 10^{-26}\,$kg,
the well-depth is $\varepsilon = 4.93\times 10^{-22}\,$J,
and the equilibrium separation is $r_\mathrm{e}=3.13\times10^{-10}\,$m.
These parameters are appropriate for neon.\cite{Sciver12}

%
\section{Results}
\setcounter{equation}{0} \setcounter{subsubsection}{0}
%

\begin{figure}[t!]
\centerline{ \resizebox{8cm}{!}{ \includegraphics*{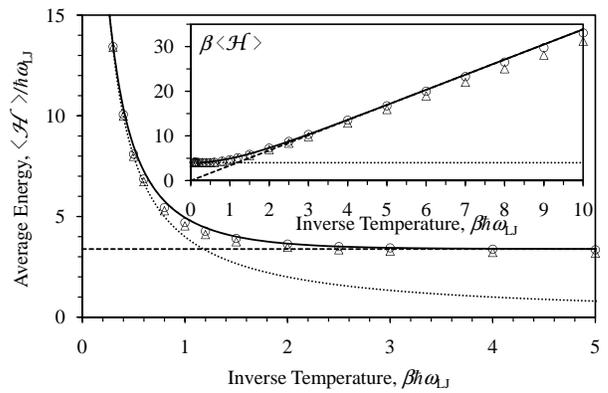} } }
\caption{\label{Fig:EvsB}
Average energy versus inverse temperature
($N=4$, $\Delta_q = r_\mathrm{e}$,
$\lambda = \kappa = m\omega_\mathrm{LJ}^2$ and $d_\mathrm{m}=0$).
The solid curve is the exact result
using $l_\mathrm{max} = 10,000$ energy levels.
The triangles are the singlet
and the circles are the pair mean field
Monte Carlo simulations.
The statistical error is less than the size of the symbols.
The dotted curve is the classical result,
$\langle E \rangle_\mathrm{cl}=N/\beta$.
The dashed line is the ground state, $E_1 = \sum_{n=1}^N \hbar \omega_n/2$.
{\bf Inset.} Average energy divided by temperature,
$\beta \langle \hat{\cal H} \rangle$, versus inverse temperature.
 }
\end{figure}

Figure~\ref{Fig:EvsB}
shows the average energy as a function of inverse temperature
using exact calculations and classical phase space simulations.
At low temperatures the phonon modes are in the ground state,
$E_1 = \sum_{n=1}^N \hbar \omega_n/2$,
and at high temperatures the system approaches the classical result,
$\langle {\cal H} \rangle_\mathrm{cl}=N/\beta$.
It can be seen in the main part of the figure
that both the singlet and pair mean field
approximations are in relatively good agreement with these limiting results
and with the exact calculations over the whole temperature regime shown.
The commutation function provides a significant correction
to the classical results at high temperatures.
In the regime of Fig.~\ref{Fig:EvsB},
symmetrization effects are negligible
and only the $d_\mathrm{m}=0$ calculations are shown.

There are two approximations in the exact calculations:
the number of energy levels used
and the domain and spacing of the grid used for the numerical quadrature.
(I persist in calling these results `exact'
because they use explicit analytic expressions
for the energy eigenvalues and eigenfunctions.)
The quadrature affects only the average density profile
without symmetrization,
and also the  average energy
and the average density profile with symmetrization.
Hence the exact calculations of the average energy
in the absence of symmetrization effects, $d_\mathrm{m}=0$,
as in Fig.~\ref{Fig:EvsB},
are  approximate only as regards to the number of energy levels
that are retained.

The exact calculations  in Fig.~\ref{Fig:EvsB} use 10,000 energy levels.
These are adequate for low and intermediate temperatures,
$\beta \hbar \omega_\mathrm{LJ} \agt 0.24$,
judged in part by comparison with results using 5,000 energy levels.
The exact data  begins to underestimate the classical result
for higher temperatures than this,
and are not shown  in Fig.~\ref{Fig:EvsB}.
One might speculate that the exact quantum result
for the average energy of the harmonic crystal
should approach the classical limit from above.
Using fewer energy levels reduces the domain of inverse temperatures
in which the exact calculations are reliable.

Both the singlet and pair mean field simulations
are practically exact at higher temperatures,
 $\beta \hbar \omega_\mathrm{LJ} \alt 0.5$ in this case.
As the temperature is decreased, the singlet mean field energy
lies between the exact energy and the classical energy.
The pair mean field result lies between the singlet mean field energy
and the exact energy.
It can be seen that at low temperatures the classical energy
is substantially less than the exact ground state energy,
but the pair mean field energy is only slightly less
than the exact ground state energy.
It may be concluded that the mean field approach
is better than a high temperature expansion
in that it yields the dominant quantum correction to the classical result
over the entire range of temperatures.

The inset of Fig.~\ref{Fig:EvsB} scales the average energy
by the inverse temperature and focusses on the low temperature regime.
For inverse temperatures $\beta \hbar \omega_\mathrm{LJ} \agt 3$,
the exact results with 10,000 energy levels
are practically indistinguishable from the ground state energy.
At low temperatures, both mean field classical phase space approximations
give a lower energy than the ground state.
For example, in the case of Fig.~\ref{Fig:EvsB},
the ground state energy is
$E_1 /\hbar \omega_\mathrm{LJ}
= 3.385 $.
At $\beta \hbar \omega_\mathrm{LJ} = 10$,
the singlet mean field theory gives
$\langle \hat {\cal H} \rangle/\hbar \omega_\mathrm{LJ} = 3.116 \pm .002$,
and the pair mean field theory gives $3.305 \pm .001 $.
At this temperature the classical result given by the simulation was
$0.4002 \pm .0002$,
which is rather close to the exact classical result of 0.4.
Here and throughout, the statistical error quoted for the simulations
is twice the standard error on the mean,
which is the 96\% confidence level.

One can conclude from the data that
at low temperatures such that the system is close to the ground state,
the mean field approximations remain viable.
The pair mean field approach substantially reduces the error
in the singlet mean field approach.
In the absence of exact data,
the difference between the pair and singlet mean field results
would give a  guide
to the quantitative accuracy of the former.

It is worth mentioning that at each temperature
the classical phase space simulations took about five minutes
on a desktop  personal computer to obtain the quoted accuracy.
In comparison,
it took about 2 days to obtain the exact results
with these energy levels and quadrature grid,
the latter being the time limiting part of the computations.
(This is independent of how many temperature points are saved.)

\begin{figure}[t!]
\centerline{ \resizebox{8cm}{!}{ \includegraphics*{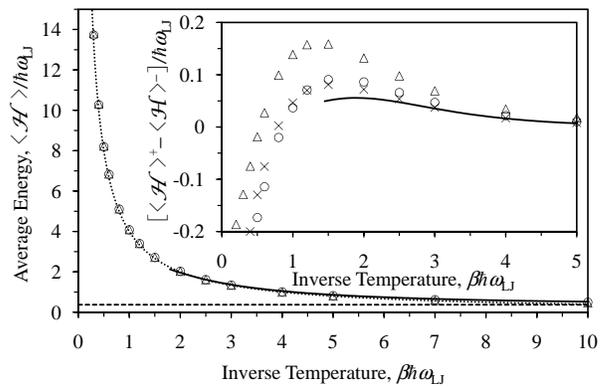} } }
\caption{\label{Fig:EvsB02}
Average energy versus inverse temperature
($N=4$, $\Delta_q = r_\mathrm{e}$,
$\kappa  = 0$, $ \lambda/ m\omega_\mathrm{LJ}^2 = 0.02$,
$d_\mathrm{m}=0$).
The solid curve is the exact result
using $l_\mathrm{max} = 20,000$ energy levels.
The triangles are the singlet
and the circles are the pair mean field
Monte Carlo simulations, respectively.
The dotted curve is the classical result,
$\langle E \rangle_\mathrm{cl}=N/\beta$,
and the dashed line is the ground state energy.
{\bf Inset.} The boson energy minus the fermion energy
using $d_\mathrm{m} \le 2$.
The crosses are Monte Carlo simulations
with the classical commutation function, $W=1$.
 }
\end{figure}

Figure~\ref{Fig:EvsB02}
shows the average energy for a weakly coupled crystal.
There is no singlet potential,
$\kappa = 0$, and the nearest neighbor spring constant
has been decreased by a factor of 50,
$ \lambda/ m\omega_\mathrm{LJ}^2 = 0.02$.
The lattice spacing and spring length is unchanged.
The ground state energy is $E_1 = 0.3757$,
which is approximately one tenth that
for the parameters of Fig.~\ref{Fig:EvsB}.


In the main body of Fig.~\ref{Fig:EvsB02}
the results of both mean field approximations
appear indistinguishable from the exact results.
At $\beta \hbar \omega_\mathrm{LJ} = 3$,
the exact energy is 1.3707 (for $l_\mathrm{max} = 20,000$; 1.3627 for 10,000),
the singlet mean field gives $1.3322 \pm .0009$
and the pair mean field gives $ 1.3443\pm .0008$.
The exact classical result here is 4/3,
compared to $1.3329 \pm .0007$ given by the simulations.
These results are for $d=0$,
so no symmetrization effects are included.
For inverse temperatures $\beta \hbar \omega_\mathrm{LJ} \alt 1.8$,
the mean field algorithms appear more reliable
than the exact results with  $l_\mathrm{max} = 20,000$.

The inset of  Fig.~\ref{Fig:EvsB02}
shows the average energy for bosons less that for fermions,
$[ \langle \hat{\cal H} \rangle^+
- \langle \hat{\cal H} \rangle^- ]/\hbar \omega_\mathrm{LJ} $,
$d_\mathrm{m} \le 2$.
Recall that for these spinless particles,
`bosons' means the fully symmetrized spatial energy eigenfunctions,
and `fermions' means the fully anti-symmetrized spatial energy eigenfunctions.
At lower temperatures, $\beta \hbar \omega_\mathrm{LJ} \agt 1.5$,
the difference is positive,
which means that
the energy for bosons is greater than that for fermions.
It can be seen that the peak difference,
which occurs at $\beta \hbar \omega_\mathrm{LJ} = 1.9$,
is about 3\% of the actual energy (exact results).
(The error in the numerical quadrature used for the exact result
is  2\% at this temperature for the energy with $d_\mathrm{m}=0$.
Hopefully for $d_\mathrm{m} \le 2$
this error is the same for bosons as for fermions
and therefore cancels.)
The classical phase space results may be described as qualitatively correct
and perhaps semi-quantitative in accuracy.
For $\beta \hbar \omega_\mathrm{LJ} \agt 2$,
the singlet mean field approximation overestimates the energy difference,
whereas the pair mean field approximation perhaps halves the error.
The classical  results, with commutation function $W=1$,
performs surprisingly well in this low temperature regime.
At higher temperatures than this all four approaches
indicate that the energy difference turns negative.
The extent to which the singlet and pair mean field predictions
agree with each other gives an indication
of their reliability in this regime.
Although the exact results are terminated at the estimated limit
of reliability of the energy,
one should note that the results in the inset of  Fig.~\ref{Fig:EvsB02}
represent the difference between two relatively large terms,
and the effects of any errors or approximation are accordingly magnified.

In Ref.~\onlinecite{Attard19},
the non-monotonic behavior of the energy difference
was attributed to two competing effects:
on the one hand  the thermal wavelength increases
with decreasing temperature,
and on the other the particles become more confined to
their lattice positions as the temperature decreases,
which reduces the amount of overlapping wave function
and non-zero symmetrization exchange.

\begin{figure}[t!]
\centerline{ \resizebox{8cm}{!}{ \includegraphics*{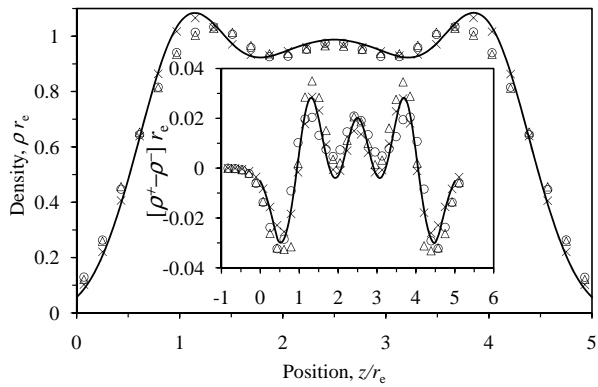} } }
\caption{\label{Fig:rhovsz02}
Density profile at $\beta \hbar \omega_\mathrm{LJ} = 2$
($N=4$, $\Delta_q = r_\mathrm{e}$,
$\kappa  = 0$, $ \lambda/ m\omega_\mathrm{LJ}^2 = 0.02$,
$d_\mathrm{m}=0$).
The solid curve is the exact result
using $l_\mathrm{max} = 20,000$ energy levels,
and the symbols are the classical phase space Monte Carlo simulations,
with the crosses being the classical result, $W=1$,
the triangles using the singlet
and the circles using the pair mean field commutation function.
{\bf Inset.} The boson density minus fermion density
using  $d_\mathrm{m} \le 2$.
 }
\end{figure}

Figure \ref{Fig:rhovsz02} shows the density profile
for this weak coupling case at $\beta \hbar \omega_\mathrm{LJ} = 2$.
The density peaks are rather broad,
with the central two particles merging into a single peak.
Interestingly enough, the density profile
spills over beyond the wall particles at $q_0 = 0$ and $q_5 = 5r_\mathrm{e}$.
There is good agreement between all four methods,
with the classical phase space simulations being closer to the exact results
at the shoulders of the density profile.

The inset of the figure shows the difference between the density of bosons
and that for fermions
with  symmetrization effects
accounted for by only nearest neighbor transpositions (dimers).
The phase space simulations may again be described as quantitatively correct.
Evidently the bulk of the symmetrization effects
are captured using the classical commutation function, $W=1$.

\begin{figure}[t!]
\centerline{ \resizebox{8cm}{!}{ \includegraphics*{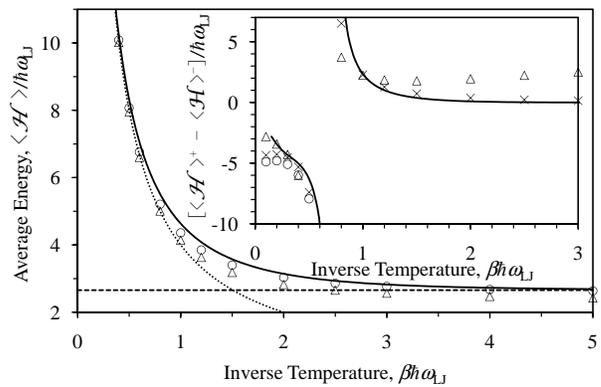} } }
\caption{\label{Fig:HvsB.1}
Average energy versus inverse temperature
($N=4$, $\Delta_q = r_\mathrm{e}/10$,
$\kappa  = 0$, $ \lambda = m\omega_\mathrm{LJ}^2 $,
$d_\mathrm{m}=0$).
The solid curve is the exact result
using $l_\mathrm{max} = 20,000$ energy levels.
The triangles are the singlet
and the circles are the pair mean field
Monte Carlo simulations, respectively.
The dotted curve is the classical result,
$\langle E \rangle_\mathrm{cl}=N/\beta$,
and the dashed line is the ground state energy.
{\bf Inset.} The boson energy minus the fermion energy
using $d_\mathrm{m} \le 2$.
The crosses are simulations with the classical commutation function, $W=1$.
}
\end{figure}

Figure~\ref{Fig:HvsB.1} is for a high density case
with lattice spacing  $\Delta_q = r_\mathrm{e}/10$.
It can be seen that the exact energy approaches the classical energy
from above as the temperature is increased.
Without symmetrization effects (main part of figure)
the mean field simulations
lie between the exact and the classical results,
with the pair mean field results  lying closer
to the exact results than the singlet mean field results
across the temperature range shown.

Both mean field results lie below
the ground state energy at low temperatures.
For example,
in this case the ground state energy is
$E_1/\hbar \omega_\mathrm{LJ} = 2.6569 $,
and at $\beta \hbar \omega_\mathrm{LJ} = 10$,
the singlet mean field gives
$\langle \hat{\cal H} \rangle/\hbar \omega_\mathrm{LJ}  =  2.336 \pm .001$,
and the pair mean field gives $2.575 \pm .001$.
Both are substantially more accurate than the classical simulation result
of $0.4001 \pm   0.0002$.

The inset to Fig.~\ref{Fig:HvsB.1}
shows the energy for bosons less that for fermions,
calculated by including only nearest neighbor transpositions,
$d_\mathrm{m} = 2$.
It can be seen that there is a pole in the exact results for fermions
at $\beta \hbar \omega_\mathrm{LJ} \approx 0.72$.
(This pole disappears when two nearest neighbor dimer transpositions
or a cyclic permutation of three consecutive particles,
$d_\mathrm{m} = 4$, are included.)\cite{Attard19}
It can be seen that the classical, $W=1$,
singlet mean field, and pair mean field commutation functions
all give this pole for $d_\mathrm{m} \le 2$ at about the same location.
There is little to choose between the three at high temperatures;
the apparent agreement of the mean field approximations with
the exact results for  $\beta \hbar \omega_\mathrm{LJ} \le 0.3$
should not be taken seriously because this is about the limit of reliability
of the exact results with $l_\mathrm{max} = 20,000$ in this case.
At intermediate and low temperatures,
$\beta \hbar \omega_\mathrm{LJ} \agt 1$,
the classical results lie closer to the exact results
than do the mean field results.
It is difficult to obtain the results for fermions accurately
when the denominator passes through zero.
In such a regime the neglected higher order terms in the expansion
contribute significantly

\begin{figure}[t!]
\centerline{ \resizebox{8cm}{!}{ \includegraphics*{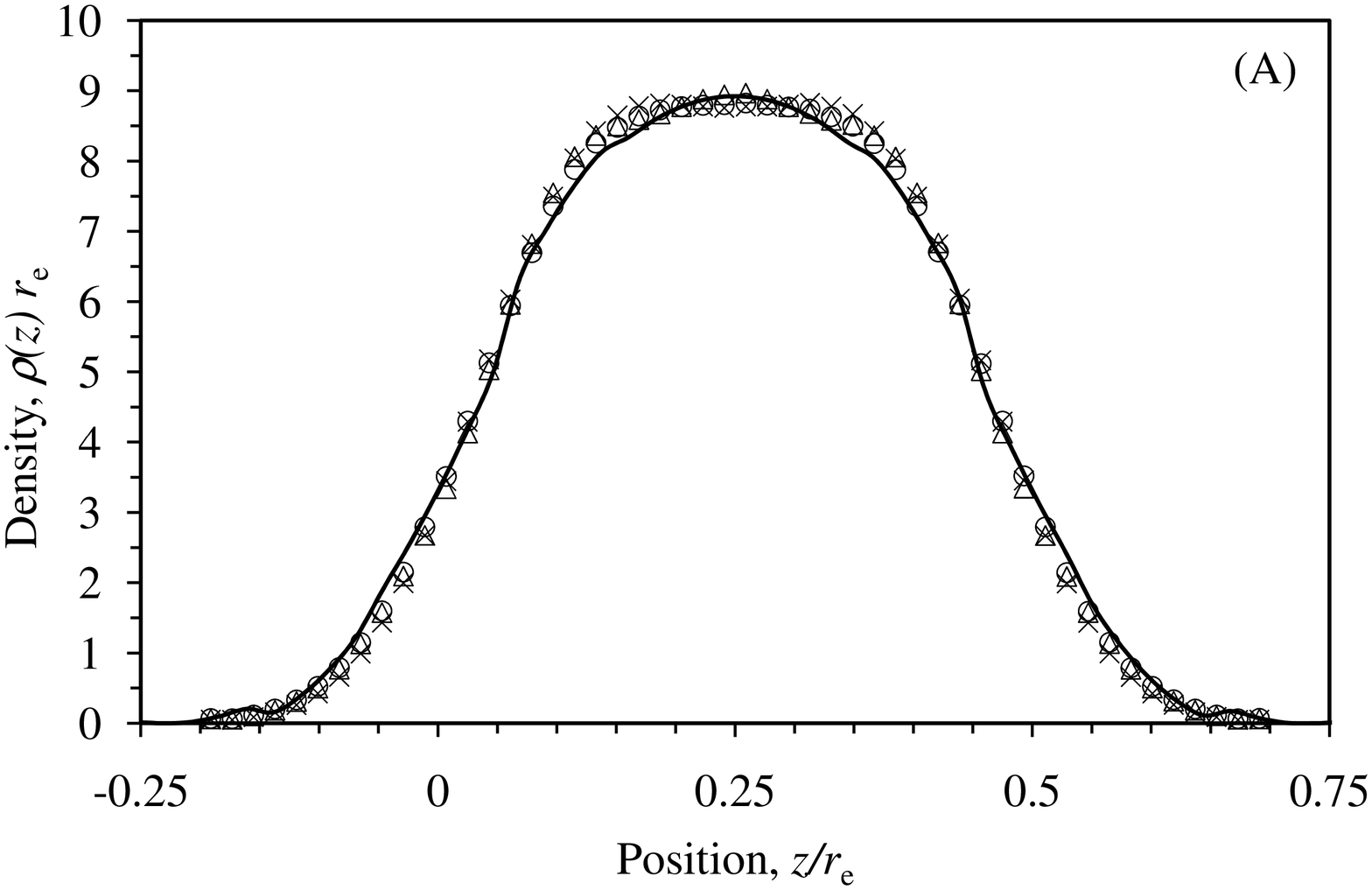} } }
\centerline{ \resizebox{8cm}{!}{ \includegraphics*{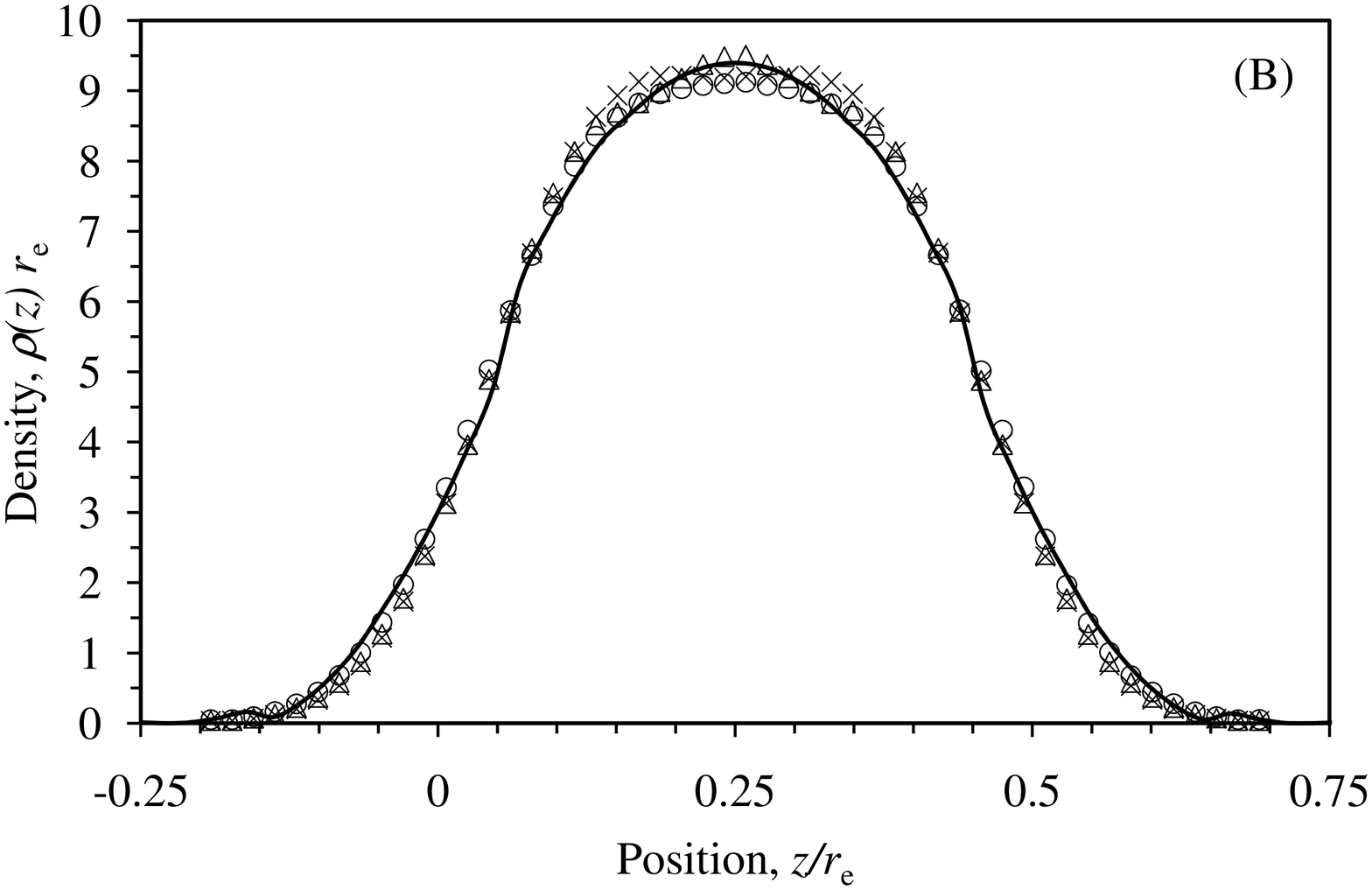} } }
\centerline{ \resizebox{8cm}{!}{ \includegraphics*{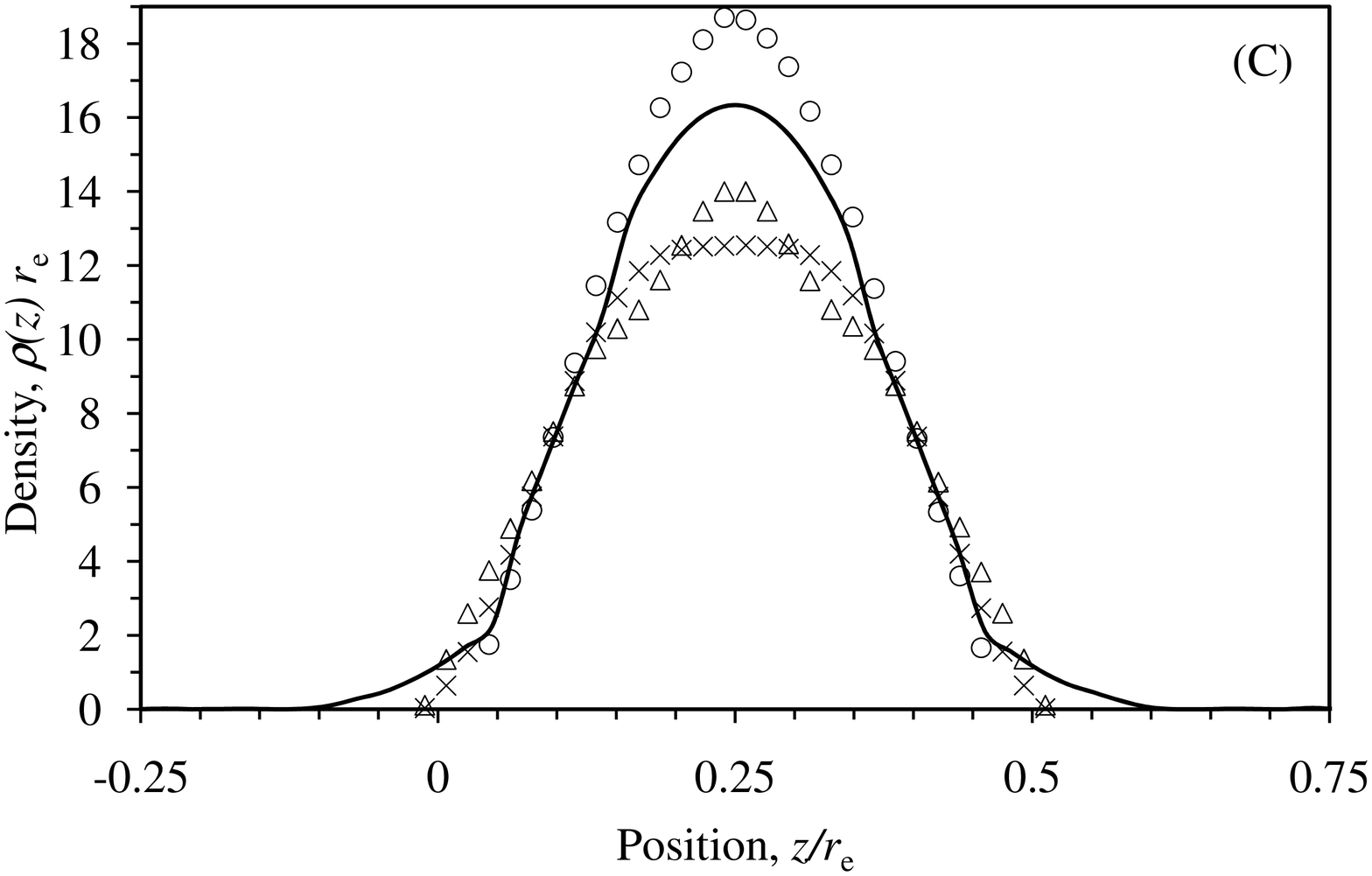} } }
\caption{\label{Fig:rvsz}
Density profile for $\beta \hbar \omega_\mathrm{LJ} =1$
($N=4$, $\Delta_q = r_\mathrm{e}/10$,
$\kappa  = 0$, $ \lambda =m\omega_\mathrm{LJ}^2$).
The solid curve is the exact result
using $l_\mathrm{max} = 20,000$ energy levels.
The crosses are the classical, $W=1$,
the triangles are the singlet,
and the circles are the pair mean field
Monte Carlo simulations, respectively.
\textbf{(A)} $d_\mathrm{m}=0$.
\textbf{(B)} Bosons, $d_\mathrm{m} \le 2$.
\textbf{(C)} Fermions, $d_\mathrm{m} \le 2$.
 }
\end{figure}

Figure~\ref{Fig:rvsz}
shows the density profile
in this high density case at $\beta \hbar \omega_\mathrm{LJ} =1$.
It can be seen that there is essentially a single density peak
in the center of the system,
and that the density spills beyond the wall particles at
$q_0 =0$ and $q_5 = 0.5$.
There is little to choose between the three simulation algorithms,
at least in the case of monomers (ie.\ no symmetrization effects,
$d_\mathrm{m}=0$).
Compared to the exact calculations
all three simulation algorithms are slightly broader at the peak.

For the case of bosons, Fig.~\ref{Fig:rvsz}B,
including nearest neighbor transpositions makes
the profile slightly narrower and more sharply peaked.
There is again good agreement between the three simulation algorithms,
with the pair mean field approach slightly underestimating
the height of the density peak.
For the case of fermions, Fig.~\ref{Fig:rvsz}C,
the exact calculations give a single peak,
that is narrower and higher than for bosons.
There is no dip in the center of this peak,
as one might have expected from Fermi repulsion.
(At the lower temperature of $\beta \hbar \omega_\mathrm{LJ} =2$,
the pair mean field profile shows a bifurcated peak,
but the other methods show a single peak similar to here).
The classical and the singlet mean field approaches
underestimate the height of the peak,
with the latter being closer to the exact results than the former.
The pair mean field approach overestimates
the height of the peak.
All three simulation algorithms miss the broad base
to the density profile at the walls
that is given by the exact calculations.
Arguably for reliable results
for fermions at this density and temperature,
one should go beyond  single dimer transpositions.

%
\section{Conclusion}
\setcounter{equation}{0} \setcounter{subsubsection}{0}
%

This paper has been concerned with ascertaining the accuracy
of a quantum Monte Carlo algorithm for an interacting system.
The algorithm is based on a formulation
of quantum statistical mechanics in classical phase space
and it uses a mean field approximation for the commutation function.
For the tests reliable exact results were required as benchmarks,
and these were obtained for a one-dimensional harmonic crystal
for which the energy eigenvalues and eigenstates can be expressed
analytically in closed form.
The analytic nature of the exact results for this model
allowed up to 20,000 energy levels to be used to establish the benchmark
results for the tests.
Previous tests of the mean field classical phase space formulation
of quantum statistical mechanics\cite{Attard18c}
used  benchmarks established for a one-dimensional
Lennard-Jones model with 50 energy eigenvalues obtained numerically.
\cite{Hernando13}   
Here it was found that the number of energy levels has a significant effect
on the statistical average at higher temperatures,
and so the present benchmarks can be relied upon in this regime.

Two versions of the mean field approximation were tested:
the singlet version, which has previously been published,
\cite{Attard18b,Attard18c}
and a pair version, which is new here.
It was found that the singlet mean field approach
was qualitatively correct over the whole temperature
(and density and coupling) regime studied.
It appeared to be exact in the high temperature limit,
and it was generally better than 10\% accurate
in the low temperature regime in which the system was
predominately in the quantum ground state.
The pair mean field algorithm significantly improved
the accuracy of the singlet algorithm
in the intermediate and low temperature regime.
In the absence of benchmark results,
the difference between the singlet and pair mean field predictions
can be used as a guide to the quantitative accuracy of the latter.

There appear to be at least two advantages
to the present mean field treatment of the commutation function
compared to evaluating it from high temperature expansions.
\cite{Wigner32,Kirkwood33,STD2,Attard18b}
First, the mean field expressions remain accurate
across the entire temperature regime,
including the ground state.
Second, algebraically the mean field expressions are relatively simple,
and computationally they are easy to implement and efficient to evaluate.
In contrast the high temperature expansions rapidly become algebraically
complex as higher order terms are included,\cite{STD2,Attard18b}
and there are corresponding challenges in their computational implementation
and numerical evaluation.

The present paper also explored wave function symmetrization effects.
This was at the dimer level,
which means the transposition of nearest neighbor particles.
It was found, somewhat surprisingly,
that combining classical Monte Carlo in classical phase space
with the symmetrization function (ie.\ neglecting the commutation function)
in some, but not all, cases
gave as good results as those obtained retaining the mean field
commutation function.
At the highest density studied,
some features of the symmetrized system
were not captured entirely by the present phase space simulations,
particularly in the case of fermions.
This suggests that retaining further terms
in the symmetrization loop expansion
(eg.\ double dimer and trimer)
may be necessary in some cases.
It may also be worth reflecting on the underlying philosophy
of the mean field approach in the presence of wave function symmetrization.

Finally, a rather interesting conclusion
from the present and earlier results\cite{Attard18c}
is that the classical component dominates,
not just in the high temperature limit
but even in the quantum ground state (for structure, not energy).
Of course quantum effects are not entirely absent,
and when present these are captured
by the present mean field commutation
function and also the symmetrization function,
but it is clear that these truly are a perturbation
on the classical prediction.
This underscores the utility of treating real world condensed matter
systems via quantum statistical mechanics formulated
as an integral over classical phase space,
rather than formulating it as a sum over quantum states,
or by parameterizing the wave function.




\end{document}